\documentclass[aps,prd,amssymb,eqsecnum,showpacs]{revtex4}
\usepackage{graphicx}
\usepackage{psfrag}

\begin{document}

\title{The Well-posedness of the  Null-Timelike Boundary Problem for Quasilinear Waves}

\author{H-O. Kreiss${}^{1,2}$   and J. Winicour${}^{2,3}$
       }
\affiliation{
${}^{1}$ NADA, Royal Institute of Technology, 10044 Stockholm, Sweden\\
${}^{2}$ Max-Planck-Institut f\" ur
         Gravitationsphysik, Albert-Einstein-Institut, 
	  14476 Golm, Germany\\
${}^{3}$ Department of Physics and Astronomy \\
         University of Pittsburgh, Pittsburgh, PA 15260, USA 
	 }

\begin{abstract}

The null-timelike initial-boundary value problem for a hyperbolic system of
equations consists of the evolution of data given on an initial characteristic
surface and on a timelike worldtube to produce a solution in the exterior of
the worldtube. We establish the well-posedness of this problem for the
evolution of a quasilinear scalar wave by means of energy estimates. The
treatment is given in characteristic coordinates and thus provides a guide for
developing stable finite difference algorithms. A new technique underlying the
approach has potential application to other characteristic initial-boundary
value problems.

\end{abstract}


\maketitle

\section{Introduction}

The use of null hypersurfaces as coordinates to describe gravitational waves,
as introduced by Bondi~\cite{bondi}, was key to the understanding and
geometric treatment of gravitational waves in the full nonlinear context of
general relativity~\cite{sachs,penrose}. In one version of the associated
characteristic initial-boundary value problem for Einstein's equations,
boundary data is given on a timelike worldtube and on an initial outgoing null
hypersurface~\cite{tam}. The physical picture underlying this null-timelike
problem is that the worldtube data represent the outgoing gravitational
radiation emanating from interior matter sources, while ingoing radiation
incident on the system is represented by the initial null data. This problem
has been developed into a Cauchy-characteristic matching scheme in which the
worldtube data is supplied by a Cauchy evolution of the interior
sources~\cite{vishu}.  See~\cite{lr} for a review. Cauchy-characteristic
matching has been implemented as a numerical evolution code in which the Bondi
news function describing the radiation is calculated at future null infinity
using a finite numerical grid obtained by Penrose
compactification~\cite{penrose}. Although characteristic evolution codes have
successfully simulated many null-timelike problems~\cite{lr} and  have
recently been applied to extract the radiation from the inspiral and merger of
a binary black hole~\cite{reis}, the well-posedness of the null-timelike
problem for the Einstein equations has not yet been established. The
characteristic formulation of the Einstein equations implies that certain
variables associated with the radiation satisfy a wave equation. Consequently,
a necessary condition for the well-posedness of the gravitational problem is
that  the corresponding problem for the quasilinear wave equation be
well-posed. In this paper, as a first step toward treating the gravitational
case, we show that the quasilinear null-timelike problem for a scalar wave
propagating on a curved space background is well posed.

The characteristic initial value problem did not receive much attention before
its importance in general relativity was recognized.  Historically, the
development of computational physics has focused on hydrodynamics, where the
characteristics typically do not define useful coordinate surfaces and there
is no generic outer boundary behavior comparable to null infinity. The
simplest problem for which the characteristic approach is useful is the
Minkowski space wave equation, which is satisfied by the components of the
fundamental special relativistic fields. Progress on the null-timelike problem
traces back to Duff~\cite{duff}, where existence and uniqueness was shown for
the linear wave equation with analytic coefficients and analytic data.
Existence and uniqueness was later extended to the $C^\infty$ case of the
linear wave equation on an asymptotically flat curved space background by
Friedlander~\cite{fried,fried2}.

The demonstration of well-posedness of the quasilinear boundary problem, i.e.
the continuous dependence of the solution on the data, depends upon
establishing estimates on the derivatives for the linearized problem. This
requires considering generic lower differential order terms~\cite{klor}.
Well-posedness depends crucially on the stability of the  problem against such
lower order perturbations. Otherwise, one cannot localize the problem and use
the principle of frozen coefficients.  

Partial results estimating the derivatives for  characteristic boundary
problems were first obtained by M{\" u}ller zum Hagen and Seifert~\cite{zum}.
Later Balean carried out a comprehensive study of the differentiability of
solutions of the null-timelike problem for the flat space wave
equation~\cite{bal1,bal2}.  He was able to establish estimates for the
derivatives tangential to the outgoing null cones but weaker estimates for the
time derivatives transverse to the cones had to be obtained from a direct
integration of the wave equation. The derivatives  tangential to the null cone
were controlled by the derivatives of the data but control of the transverse
time derivative required two derivatives of the data. Balean concentrated on
the differentiability order of the solution and did not discuss the
implications for well-posedness of the quasilinear problem.

Frittelli~\cite{frit} made the first explicit consideration of well-posedness
of the null-timelike problem for the wave equation. She adopted the approach
of Duff, in which the characteristic formulation of the wave equation is
reduced to a canonical first order differential form, in close analogue to the
symmetric hyperbolic formulation of the Cauchy problem. The energy associated
with this first order reduction gives estimates for the derivatives of the
field tangential to the null hypersurfaces. As in Balean's treatment,
weaker estimates for the time derivatives were obtained indirectly so that
well-posedness is not ensured when lower order differential order terms or
source terms are included as required for the quasilinear case,
as she was careful to point out.

A difficulty underlying the problem can be illustrated in terms of the
1(spatial)-dimensional wave equation
\begin{equation}
             ( \partial_{\tilde t}^2 -\partial_{\tilde x}^2)\Phi=0,
             \label{eq:tilde1d}
\end{equation}
where $(\tilde t,\tilde x)$ are standard space-time coordinates.
The conserved energy
\begin{equation}
       \tilde  E(\tilde t)= \frac{1}{2} \int d\tilde x  \bigg( (
        \partial_{\tilde t}\Phi)^2  +(\partial_{\tilde x}\Phi)^2 \bigg )
       \label{eq:tildeE}
\end{equation}
leads to the well-posedness of the Cauchy problem. In characteristic
coordinates $(t=\tilde t -\tilde x, \, x=\tilde t +\tilde x)$,
the wave equation transforms into
\begin{equation}
   \partial_t \partial_x \Phi =0.
   \label{eq:1dphi}
\end{equation}
The conserved energy on the characteristics $t={\rm const.}$,
\begin{equation}
       \tilde   E(t) = \int dx (\partial_ x \Phi)^2,
\end{equation}
no longer controls the derivative $\partial_t \Phi$.

The first  proof of well-posedness of the characteristic initial
value problem valid for the quasilinear wave equation has been the work of
Rendall~\cite{rend}, who considered the double null problem where data is
given on a pair of intersecting characteristic hypersurfaces. Rendall did not
treat the characteristic problem head-on but reduced it to a standard Cauchy
problem with data on a spacelike hypersurface  passing through the
intersection of the characteristic hypersurfaces. Well-posedness than follows
from  the classic result for the Cauchy problem. He extended his treatment to
establish the well-posedness of the double-null formulation of the Einstein
gravitational problem. The double null problem treated by Rendall is a
limiting case of the null-timelike problem considered in this paper. However,
Rendall's approach cannot be applied to the null-timelike problem. Also, the
reduction to a Cauchy problem does not provide guidance for the development of
a stable finite-difference approximation based upon characteristic
coordinates.

Another limiting case of the null-timelike problem is the Cauchy problem on
a characteristic cone, corresponding to the limit in which the timelike worldtube
has shrunk to a nonsingular worldline. This problem is difficult to treat in
characteristic coordinates because of their singular nature at the vertex of the
cone. However, Choquet-Bruhat, Chru\'{s}ciel and Mart\'{i}n-Garc\'{i}a have been
able to establish the existence of solutions to this problem, for both the
scalar and gravitational case, by treating it in harmonic coordinates adapted
to the null cones~\cite{cbcmg1,cbcmg2}.

Here we consider the null-timelike problem for the quasi-linear wave equation
in second differential form in terms of characteristic coordinates.. The usual
technique for showing that the initial-boundary value problem for a hyperbolic
system of partial differential equations is well posed is to split the problem
into a Cauchy problem and local halfplane problems and show that these
individual problems are well posed. This works for hyperbolic systems based
upon a spacelike foliation, in which case signals propagate with finite
velocity. Besides the existence and uniqueness of a solution, well-posedness
implies that the solution depend continuously on the data with respect to an
appropriate norm.  For (\ref{eq:tilde1d}), the solutions to the Cauchy problem
with compact initial data on $\tilde t=0$ are square integrable and
well-posedness can be established using the $L_2$ norm  (\ref{eq:tildeE}).

However, In characteristic coordinates the 1-dimensional wave equation
(\ref{eq:1dphi}) admits signals traveling in the $+x$-direction with infinite
coordinate velocity. In particular, initial data of compact support
$\Phi(0,x)=f(x)$ on the characteristic $t=0$ admits the solution $\Phi = g(t)
+f(x)$, provided that $g(0)=0$.  Here $g(t)$ represents the profile of a wave
which travels from past null infinity ($x\rightarrow -\infty$) to future null
infinity  ($x\rightarrow +\infty$). Thus, without a boundary condition at past
null infinity, there is no unique solution and the Cauchy problem is ill
posed. Even with the boundary condition $\Phi(t,-\infty)=0$,  a source of
compact support $S(t,x)$ added to (\ref{eq:1dphi}), i.e.
\begin{equation}
   \partial_t \partial_x \Phi =S,
   \label{eq:1dphis}
\end{equation}
produces waves propagating to $x=+\infty$ so that although the solution is 
unique it is still not square integrable.

On the other hand, consider the modified problem obtained by
setting $\Phi=e^{ax}\Psi$,
\begin{equation} 
 \partial_t (\partial_x+a) \Psi=F \, ,  \quad \Psi(0,x)
    = e^{-ax}f(x) \,  ,\quad a>0
       \label{eq:1dpsi}
\end{equation} 
where $F=e^{-ax}  S$. With the boundary condition $\Psi(t,-\infty)=0$, the
solutions to (\ref{eq:1dpsi}) vanish at $x=+\infty$ and are square
integrable. As a result, the Cauchy problem (\ref{eq:1dpsi}) is well posed
with respect to an $L_2$ norm. For the simple example where $F=0$,
multiplication of (\ref{eq:1dpsi}) by
$(2a \Psi+\partial_x \Psi+\frac{1}{2}\partial_t \Psi)$ and integration by
parts gives
\begin{equation}
       \frac{1}{2}\partial_t  \int dx \bigg( 
             (\partial_ x \Psi)^2+2a^2 \Psi^2 \bigg)
            =\frac{a}{2} \int dx \bigg(2(\partial_ t \Psi)\partial_ x \Psi 
           -  (\partial_ t \Psi)^2 \bigg)   
         \le \frac{a}{2} \int dx  (\partial_ x \Psi)^2 .
\end{equation} 
The resulting inequality
\begin{equation}
       \partial_t  E \le {\rm const.} E 
\end{equation} 
for the energy
\begin{equation}
      E=\frac{1}{2}  \int dx \bigg( (\partial_ x \Psi)^2+2a^2 \Psi^2 \bigg)
      \label{eq:1denergy}
\end{equation} 
provides the estimates for $\partial_x \Psi$ and $\Psi$ which are necessary
for well-posedness. Estimates for $\partial_t \Psi$, and other higher
derivatives, follow from applying this approach to the derivatives of 
(\ref{eq:1dpsi}). The approach can be extended to include the source term $F$
and other generic lower differential order terms. This allows well-posedness
to be extended to the case of variable coefficients and, locally in time, to
the quasilinear case.

The 2(spatial)-dimensional model problems considered in Sec.~\ref{sec:model}
illustrate how this approach generalizes to the multi-dimensional case. We
consider the model problems in the modified form analogous to 
(\ref{eq:1dpsi}). By means of this technique, the characteristic
initial-boundary value problem can again be treated by first considering
Cauchy and half-plane problems. The demonstration of well-posedness of these
model problems presents the underlying ideas in a transparent form.

Our main technique is the use of energy estimates. Although the model problems
are treated in the modified form, the results can be translated back to the
original problem.  For example, the modification in going from  
(\ref{eq:1dphis}) to  (\ref{eq:1dpsi}) leads to an effective modification of
the standard energy for the problem.  Rewritten in terms of the original
variable $\Phi=e^{ax}\Psi$, (\ref{eq:1denergy}) corresponds to the energy 
\begin{equation}
     E=\frac{1}{2}  \int dx e^{-2ax}  \bigg( (\partial_ x \Phi)^2+a^2 \Phi^2 \bigg ).
     \label{eq:enorm}
\end{equation}
Thus while the Cauchy problem for (\ref{eq:1dpsi}) is ill posed with respect
to the $L_2$ norm it is well posed with respect to the exponentially weighted
norm (\ref{eq:enorm}). However, rather than modifying the norm, for technical
simplicity we deal with the modified  variable $\Psi$.

The general arguments presented for our model problems can be applied to a
wide range of quasilinear characteristic problems. Our motivation for the work
here is the application to the null-timelike problem for the quasilinear wave
equation for a scalar field $\Phi$ in an asymptotically flat curved space
background with source $S$,
\begin{equation}
        g^{ab}\nabla_a \nabla_b \Phi = S (\Phi,\partial_c \Phi, x^c),
        \label{eq:qw}
\end{equation}
where the metric $g^{ab}$ and its associated covariant  derivative $\nabla_a$
are explicitly prescribed functions of $(\Phi,x^c)$.

The corresponding flat space wave equation,
\begin{equation}
   (-\partial_{\tilde t}^2 + \partial _{\tilde x}^2 + \partial _{\tilde y}^2
     + \partial _{\tilde z}^2)\Phi =  S,
\end{equation}
takes the form
\begin{equation}
   {1\over r} (-2  \partial_u  \partial_r +\partial_r^2) (r\Phi)+
   {1\over r^2\sin \theta}\partial_\theta(\sin \theta\partial_\theta \Phi)
       +  {1\over r^2 \sin^2 \theta} \partial_\phi^2 \Phi=S
   \label{eq:fwe}
\end{equation} 
in null-spherical coordinates $(u,r,\theta,\phi)$ consisting of
a retarded time $u={\tilde t}-r$ and standard spherical coordinates
$(r,\theta,\phi)$. In these coordinates, the  Minkowski metric is
\begin{equation}
   ds^2 =-du^2-2du dr  +r^2(d\theta^2+\sin^2 \theta d\phi^2).
   \label{eq:minkn}
\end{equation}
The null-timelike problem consists of determining $\Phi$ in the region
$(r>R,u>0)$ given data $\Phi(u,R,\theta,\phi)$ on the timelike worldtube $r=R$
and  $\Phi(0,r,\theta,\phi)$ on the initial null hypersurface $u=0$.

In an asymptotically flat background, the metric (\ref{eq:minkn})
generalizes to the Bondi-Sachs form
\begin{equation}
   g_{ab}dx^a dx^b = -(e^{2\beta}W-r^{-2}h_{AB}W^A W^B)du^2 -2e^{2\beta}dudr
      -2h_{AB}W^B dudx^A   +r^2h_{AB}dx^A dx^B,
      \label{eq:nullmet}
\end{equation}
where $x^A$ are angular coordinates such that $(u,x^A)={\rm const.}$ along the
outgoing null rays. Here the radial coordinate $r$ is a surface area
coordinate so that the  area of the topological spheres $(u,r)={\rm const.}$
is $4\pi$ as measured by the conformal 2-metric $h_{AB}$.
In the curved space version of angular
coordinates analogous to (\ref{eq:minkn}), $\det(h_{AB}) =\sin^2\theta$.

In Sec.~\ref{sec:af}, we treat the null-timelike problem for the quasilinear
wave equation (\ref{eq:qw}) with asymptotically flat Lorentzian metric (\ref{eq:nullmet}),
\begin{eqnarray}
     \label{eq:3dwe}
  {1\over r} (-2  \partial_u  \partial_r +W \partial_r^2) (r\Phi)
    +(\partial_r  W)\partial_r\Phi
    & -&\frac{1}{r^2}D_A(W^A \partial_r \Phi) 
        -{1\over r^2} \partial_r (W^A D_A \Phi) 
     +{1\over r^2} D_A(e^{2\beta} D^A \Phi) \nonumber \\
    & = &e^{2\beta} S(\Phi,\partial_c \Phi,x^c) ,\\
     \Phi(0,r,x^A)=f(r,x^A)\,  ,\quad \Phi(u,R,x^A) &=&q(u,x^A)\,, \quad R\le  r<\infty  \, , \nonumber
         \quad u\ge 0.
      \end{eqnarray}    
Here
$D_A$ is the 2-dimensional covariant derivative with respect to $h_{AB}$ and
the metric coefficients $(W,\beta,W^A,h_{AB})$ depend smoothly upon
$(\Phi,u,r,x^A)$ and the source $S$ depends smoothly upon  $(\Phi,\partial_a
\Phi,u,r,x^A)$.

An essential part of any initial-boundary value problem is the compatibility
between the data at the intersection between the initial hypersurface and the
boundary, i.e. at $(u=0,r=R)$ in the above case. This compatibility affects
the differentiability of the resulting solution. In order to avoid difficult
issues of analysis, we only give a rigorous treatment for the case of smooth
initial and boundary data with compact support bounded away from the
intersection, in which case the solution is $C^\infty$ locally in time. See
the work of Balean~\cite{bal1,bal2} for a discussion of the differentiability
of the solution in the general case.

We assume that as $r\rightarrow \infty$ (the approach to null infinity) that
the problem reduces to the flat space problem (\ref{eq:fwe}), so that the
coefficients have the asymptotic behavior $W=1+O(1/r)$, $\beta=0+O(1/r)$,
$W^A=O(1)$ and $h_{AB}=q_{AB}+O(1/r)$, where $q_{AB}$ is the unit sphere
metric. The results of Friedlander~\cite{fried2} then imply that the scalar
wave falls off as $\Phi\sim \Phi_0(u,x^A)/r$ where $ \Phi_0$ is the asymptotic
radiation field. 

\bigskip

In  Sec.~\ref{sec:af}, we establish our  {\bf Main Theorem:}

\medskip

\noindent The nullcone-worldtube problem
(\ref{eq:3dwe}) is well posed for smooth, compatible initial data
$f(r,x^A)$ and boundary data $q(u,x^A)$ subject to the conditions
that $f=O(r^{-1})$, $S=O(r^{-3})$ and a positivity condition that
the principal part of the wave operator reduces to
an elliptic operator in the stationary case. 

\bigskip

Our treatment is based upon energy
estimates obtained by integration by parts with respect to the characteristic
coordinates. As a result, the analogous finite difference estimates obtained
by {\em summation by parts} provide guidance for the development of a stable
numerical evolution algorithm for (\ref{eq:3dwe}).

\section{Well-posedness of model characteristic problems}
\label{sec:model}

We consider here several model 2(spatial)-dimensional problems which reveal
the essential features underlying a well posed characteristic initial-boundary
value problem. For simplicity of notation, we indicate partial derivatives by
subscripts, e.g. $\Phi_t(t,x,y)=\partial_t \Phi(t,x,y)$. Also, we denote the
$L_2$ scalar product and norm over the ${x,y}$ domain by  $(\Phi_1,\Phi_2)$
and $\|\Phi\|^2=(\Phi,\Phi)$.

We consider model linear problems with constant coefficients but show that the
problems are stable against lower order perturbations. We also obtain
estimates for arbitrarily high derivatives. Thus we can use standard
techniques~\cite{klor} to establish the well-posedness of the corresponding
problem with smooth variable coefficients. For the extension to the
quasilinear case, we require that the coefficients depend smoothly upon the
field $\Phi$ with nonsingular behavior in the neighborhood of the initial
data. Then well-posedness, locally in time, of the quasilinear problem also follows
from standard techniques~\cite{klor}.  (See the Appendix of~\cite{wpe} for
details concerning how these standard techniques apply to hyperbolic systems
in second differential order form.)

Our goal is to show the well-posedness of the strip problem
$$ 2\Phi_{tx}=\left((1-x)^2\Phi_x\right)_x+\Phi_{yy}+
  b\Biggl(\left((1-x)^2\Phi_x\right)_y+ \left((1-x)^2\Phi_y\right)_x\Biggr)
$$
in the domain
                $$ 0\le x \le 1,\quad -\infty < y < \infty,\quad  t\ge 0 $$
with initial and  boundary conditions
$$ \Phi(0,x,y)=f(x,y),\quad  \Phi(t,0,y)=q(t,y),$$
respectively. The method used to show that this problem is well posed
applies to the compactified version of the null-timelike boundary problem for 
the wave equation (\ref{eq:3dwe}) treated in Sec.~\ref{sec:af}. As explained
in the Introduction, we treat the problem in the modified form obtained by the
change of variable $\Phi=e^{ax} \Psi$, $a>0$.

\subsection{The Cauchy problem}
\medskip
We first consider the Cauchy problem
\begin{eqnarray}
        (\Psi_x+a\Psi)_t&= \Psi_{yy}-2b\Psi_y,
           \quad  {\bar x}=(x,y)\in R^2,\quad t\ge 0,\\
          & \Psi(0,x,y)=f(x,y),
\label{eq:1.1}
\end{eqnarray}
where $x$ and $t$ are both characteristic coordinates. Here $a,~b$ are real
constants and $f(x,y)\in C_0^\infty$ (a smooth function with compact support).
As explained in the Introduction, we investigate the behavior of square
integrable solutions, so that $\Psi(t,\pm \infty,y)=0$.

\subsubsection{The Fourier method}
We first solve the problem by Fourier transform. Let 
$$
 \hat f({\bar \omega})=\frac{1}{2\pi}\int_{R^2} e^{-i{\bar \omega}\cdot
      {\bar x}} f({\bar x}) dx dy,
   \quad {\bar \omega}=(\omega_1,\omega_2) \, {\rm real,}
$$
denote the Fourier transform of $f$ and $\hat \Psi(t,\bar \omega)$ the Fourier
transform of $\Psi$ Then $\hat \Psi(t,\bar \omega)$ is the solution of
\begin{eqnarray}
 (i\omega_1 +a)\hat \Psi_t&=-\left(\omega_2^2+2bi\omega_2\right)\hat \Psi,\\
   &\hat \Psi(0,\bar \omega)=\hat f(\bar \omega) ,
\label{eq:1.2}
\end{eqnarray} 
i.e.
$$
   \hat \Psi_t=s\hat \Psi,$$
where
\begin{eqnarray}
     s=-{\omega_2^2+2bi\omega_2\over i\omega_1+a}=
    -{(\omega_2^2+2bi\omega_2)(a-i\omega_1)\over a^2+\omega_1^2}.
\label{eq:1.3}
\end{eqnarray}
Therefore
\begin{eqnarray}
   \Re s&=-{a\omega_2^2+2b\omega_1\omega_2\over a^2+\omega_1^2}, \nonumber \\
  \Im s&={\left(\omega^2_2\omega_1-2ab\omega_1\right)\over a^2+\omega_1^2}.
\label{eq:1.4}
\end{eqnarray}

We now discuss the dependence of the solutions on $a,b$ in detail.

\bigskip

\noindent {\bf 1) $b=0,~a>0.$} By (\ref{eq:1.4}),
$$\Re s=-{a\omega_2^2\over \omega_1^2+a^2}\le 0.$$
There are no exponentially growing solutions.

\bigskip

\noindent
{\bf 2) $b=0,~a=0.$} By (\ref{eq:1.4}),
$$
    \Re s=0,\quad |{\Im}\,s|\to\infty\quad {\rm for}\quad 
   |\omega_1|\to 0,~\omega_2\ne 0.$$
Therefore the solution of (\ref{eq:1.1}) loses all smoothness in time if
$\hat f(0,\omega_2)\ne 0.$

\bigskip

\noindent
{\bf 3) $b=0,~a<0.$} By (\ref{eq:1.4}),
$$ \Re s \rightarrow +\infty \quad {\rm for}
      ~\omega_2 \rightarrow \infty.$$
Thus there is unbounded exponential growth and the problem is ill posed.

\bigskip

\noindent
{\bf 4) $b\ne 0,~a>0.$} By (\ref{eq:1.4}),
$$
    \Re s=-{a(\omega_2+{b\over a}\omega_1)^2\over \omega_1^2+a^2}
     +{b^2\omega_1^2\over a(\omega_1^2+a^2)}
     \le {b^2\over a}.
$$
There is exponential growth but the growth
is bounded independently of $\bar \omega$.
\bigskip

\noindent
{\bf 5) $b\ne 0,~a=0.$} By (\ref{eq:1.4}),
$$ \Re s=-{2b\omega_2\over \omega_1}.$$
Thus there is unbounded exponential growth as $\omega_1\rightarrow 0$.
The same is true if $a<0$.

\medskip

We now express our results in a more general setting.

{\bf Definition 2.1}. {\em We call the Cauchy problem  well posed if, for
every $f\in C_0^\infty,$ there is a unique, smooth, square integrable solution
and if there is a constant $\alpha$ which does not depend on $\bar \omega$
such that
$$ \Re s\le \alpha.$$}
The problem is ill posed if there is no upper bound $\alpha,$ i.e.,
there is a sequence $ \bar \omega^{(j)}$ such that
$$\lim _{j\to\infty} \Re s_j=\infty.$$
 
\par

\bigskip

\noindent
{\bf Theorem 2.1}. {\em The Cauchy problem (\ref{eq:1.1}) is
well posed if $a>0.$ But it is ill posed if $a<0$ or $a=0,~b\ne 0.$}

\subsubsection{The energy method}

For the generalization to variable coefficients it is necessary to show that
the differential equation (\ref{eq:1.1}) is stable against lower order
perturbations. For this purpose we first apply the energy method to the
doubly-characteristic Cauchy problem
\begin{eqnarray}
      (\Psi_x+a\Psi)_t &= \Psi_{yy}-2b\Psi_y-c\Psi_x+d\Psi_t+e\Psi +F(t,x,y),
      \label{eq:ecauch}  \\
& \Psi(0,x,y)=f(x,y)\, , \quad -\infty<x,y<\infty \, ,  \quad  t\ge 0. 
\nonumber
\end{eqnarray} Here $a>0,~ b,~c,~d,~e$ are real constants and $F$ is a forcing
(source) term of compact spatial support.

The term $d\Psi_t$ can be absorbed into the left hand side and we obtain
$\left(\Psi_x+(a-d)\Psi\right)_t$. Therefore we neglect this term and assume
that $a$ is sufficiently large so that $a-d>0$. We also neglect the term
$e\Psi$ because it has no influence on the required energy estimates..
Therefore we consider the corresponding Cauchy problem for
\begin{eqnarray}
 &(\Psi_x+a\Psi)_t=\Psi_{yy}-2b\Psi_y-c\Psi_x+F.
\label{eq:1.5}
\end{eqnarray}

We now derive an energy estimate. By (\ref{eq:1.5}),
\bigskip
$$
 (\Psi,\Psi_{xt})+a(\Psi,\Psi_t)=-(\Psi_x\Psi_t)
         +{a\over 2}\partial_t   \|\Psi\|^2
      =(\Psi,\Psi_{yy})-(\Psi,2b\Psi_y+c\Psi_x)+(\Psi,F).
$$
Since
$$
  (\Psi_x,\Psi_t)=\left({2\over\sqrt{a}}\Psi_x,{\sqrt{a}\over 2}
         \Psi_t  \right)\le
        {2\over a}\|\Psi_x\|^2+{a\over 8}\|\Psi_t\|^2,
$$
integration by parts gives
\begin{equation}
 {a\over 2}\partial_t \|\Psi\|^2+\|\Psi_y\|^2=(\Psi_x,\Psi_t)
   +(\Psi,F)\le {2\over a}\|\Psi_x\|^2+{a\over 8}\|\Psi_t\|^2
     +\frac{1}{2}(\|(\Psi\|^2+\|F\|^2).
\label{eq:1.6}
\end{equation}

Next, 
$$
  (\Psi_t,\Psi_{xt})+a\|\Psi_t\|^2=-\frac{1}{2}\partial_t \|\Psi_y\|^2
   -2b(\Psi_t,\Psi_y)-c(\Psi_t,\Psi_x) +(\Psi_t,F).
$$
Since
$$
c(\Psi_t,\Psi_x)=\left({\sqrt{a}\over 2}\Psi_t,{2c\over\sqrt{a}}\Psi_x\right)
\le {a\over 8}\|\Psi_t\|^2+{2c^2\over a}\|\Psi_x\|^2, 
$$
$$
(\Psi_t,F)\le {a\over 8}\|\Psi_t\|^2+{2\over a}\|F\|^2,
$$
$$
2b(\Psi_t,\Psi_y)=\left({\sqrt{a}\over 2}\Psi_t,{4b\over\sqrt{a}}\Psi_y\right)\le
{a\over 8}\|\Psi_t\|^2+{8b^2\over a}\|\Psi_y\|^2, 
$$
we obtain
\begin{equation}
   {5a\over 8}\|\Psi_t\|^2+\frac{1}{2}\partial_t \|\Psi_y\|^2
   \le {\rm const.}\left(\|\Psi_x\|^2 +\|\Psi_y\|^2+\|F\|^2\right).
\label{eq:1.7}
\end{equation}

Next,
$$
(\Psi_x,\Psi_{xt})+a(\Psi_x,\Psi_t)=(\Psi_x,\Psi_{yy})
   -(\Psi_x,2b\Psi_y+c\Psi_x)+(\Psi_x,F).
$$
Since $(\Psi_x,\Psi_{yy})=-(\Psi_{xy},\Psi_y)=0,$ we obtain
\begin{equation}
  \frac{1}{2} \partial_t \|\Psi_x\|^2\le {\rm const.}  
 \left(\|\Psi_x\|^2+\|\Psi_y\|^2+\|F\|^2\right)+{a\over 8}\|\Psi_t\|^2.
\label{eq:1.8}
\end{equation}

Adding (\ref{eq:1.6})--(\ref{eq:1.8}) gives the energy estimate
\begin{equation}
 {3a\over 8}\|\Psi_t\|^2+\|\Psi_y\|^2+\frac{1}{2}\partial_t \left(\|\Psi_x\|^2
  +\|\Psi_y\|^2+a\|\Psi\|^2\right)
   \le {\rm const.}
     \left(\|\Psi_x\|^2+\|\Psi_y\|^2+\|\Psi\|^2+\|F\|^2\right).
\label{eq:1.9}
\end{equation}
We have proved:
\bigskip

\noindent
{\bf Theorem 2.2}. {\em The Cauchy problem (\ref{eq:ecauch}) is well posed with respect
to the $L_2$ norm if $(a-d)>0.$ There is an energy estimate. Also, the problem is stable
against lower order perturbations. In addition, estimates for the higher derivatives of
$\Psi$ follow from the equations obtained by differentiating (\ref{eq:ecauch}) }.

\bigskip

We now consider the Cauchy problem
\begin{eqnarray}
     & (\Psi_x+a\Psi)_t = \Psi_{xx}+ \Psi_{yy} +F(t,x,y),
\label{eq:1.10}     \\
  &\Psi(0,x,y)=f(x,y), \quad -\infty<x,y<\infty \, , \quad   t\ge 0, \nonumber
\end{eqnarray}
where $x$ is a characteristic coordinate but $t$ is timelike.
We again derive an energy estimate. 

We have
\begin{eqnarray}
\left(\Psi,\Psi_{xt}+a\Psi_t\right)&
=&-(\Psi_x,\Psi_t)+{a\over 2}\partial_t \|\Psi\|^2 \nonumber \\
&=&-(\|\Psi_x\|^2+\|\Psi_y\|^2)+(\Psi,F), \nonumber
\end{eqnarray}
i.e.
\begin{equation}
{a\over 2}\partial_t \|\Psi\|^2+\|\Psi_x\|^2+\|\Psi_y\|^2=(\Psi_x,\Psi_t)+(\Psi,F)
\le{a\over 8}\|\Psi_t\|^2+{2\over a}\|\Psi_x\|^2+(\Psi,F).
\label{eq:1.12}
\end{equation}
Next,
\begin{eqnarray}
  &\left(\Psi_t,\Psi_{xt}+a\Psi_t\right)=(\Psi_t,\Psi_{xt})+a\|\Psi_t\|^2
 =a\|\Psi_t\|^2& \nonumber \\
 &=-\frac{1}{2}\partial_t (\|\Psi_x\|^2+\|\Psi_y\|^2)+(\Psi_t,F),& \nonumber
\end{eqnarray}
i.e.
\begin{equation}
 a\|\Psi_t\|^2+\frac{1}{2}\partial_t (\|\Psi_x\|^2+\|\Psi_y\|^2)=(\Psi_t,F)
    \le {a\over 8}\|\Psi_t\|^2+{2\over a}\|F\|^2.
\label{eq:1.13}
\end{equation}
Combining (\ref{eq:1.12}) and (\ref{eq:1.13}) as before, we obtain the desired estimate
\begin{equation}
 {3a\over 4}\|\Psi_t\|^2+\|\Psi_x\|^2+\|\Psi_y\|^2
     +\frac{1}{2}\partial_t  \left(\|\Psi_x\|^2
    +\|\Psi_y\|^2+a\|\Psi\|^2\right)
     \le {\rm const.}
    \left(\|\Psi_x\|^2+\|\Psi_y\|^2+\|\Psi\|^2+\|F\|^2\right).
\end{equation}
\noindent
{\bf Remark }. {\em As before, we can add a general lower order expression and
still obtain the estimate. Also, we can estimate all derivatives.}

\subsection{The half-plane problem}
\label{sec:hp}.

We now apply the energy method to the double-null halfplane problem for (\ref{eq:1.5}),
\begin{equation}
 (\Psi_x+a\Psi)_t=\Psi_{yy}+2b\Psi_y-c\Psi_x+F,\quad 0\le x<\infty,
   ~ -\infty< y <\infty,\quad t\ge 0,
\label{eq:2.1}
\end{equation}
with initial and boundary data
\begin{equation}
   \Psi(0,x,y,)=f(x,y),\quad \Psi(t,0,y)=0
\label{eq:2.2}
\end{equation}
and source $F(t,x,y)$ of compact support.

There are no difficulties to derive the basic estimate (\ref{eq:1.9}) because
for the estimates  (\ref{eq:1.6})--(\ref{eq:1.8}) we require only that $
\Psi(t,0,y)=0.$ To obtain estimates for higher derivatives we have to proceed
in the following way.

\medskip

We differentiate (\ref{eq:2.1}) with respect to $y$. Since $\Psi_y(t,0,y)=0$,
we obtain the same problem for  $\Psi_y$ and therefore we obtain estimates for
$$
   \|\Psi_{yy}\|^2,\quad \|\Psi_{xy}\|^2.
$$
If we differentiate (\ref{eq:2.1}) two times with respect to $y$, we obtain
estimates for the third derivatives. The corresponding results hold for
$t$-derivatives, e.g. 
$$
\|\Psi_{t}\|^2,\quad \|\Psi_{yt}\|^2 ,\quad \|\Psi_{xt}\|^2.
$$ 

Now we differentiate (\ref{eq:2.1}) with respect to $x.$
\begin{equation}
   (\Psi_{xx}+a\Psi_x)_t=\Psi_{yyx}+R.
\label{eq:2.3}
\end{equation}
Here $R$ consists of source terms and terms which we have already estimated.
(\ref{eq:2.3}) gives us 
\begin{equation}
 (\Psi_{xx},\Psi_{xxt})+a(\Psi_{xx},\Psi_{xt})=(\Psi_{xx},\Psi_{yyx})
   +(\Psi_{xx},R).
\label{eq:2.4}
\end{equation}
We obtain
$$
     \frac{1}{2}\partial_t  \|\Psi_{xx}\|^2\le 
 \frac{1}{2}\bigg ( (1+a^2)\|\Psi_{xx}\|^2+\|\Psi_{xt}\|^2+\|\Psi_{yyx}\|^2
     + \|R\|^2\bigg ),
$$
where we already have estimates for $\|\Psi_{xt}\|^2$ and $\|\Psi_{yyx}\|^2$. 
The process can be continued.

{\bf Remark }. {\em Inhomogeneous boundary data $\Psi(t,0,y) =q(t,y)$ may be
treated in the same way through the transformation $\Psi \to \Psi -q e^{-x}$
and absorbing the boundary data in the source term $F$. We can also treat the
timelike-null halfplane problem for (\ref{eq:1.10}) in the same way.}

\subsection{The strip problem}
\label{sec:2strip}

As a prototype of the compactified wave equation considered in
Sec.~\ref{sec:af}, we consider the strip problem
\begin{equation}
   2(\Psi_x+a\Psi)_t=\left((1-x)^2\Psi_x\right)_x+\Psi_{yy}
  +b\Biggl(\left((1-x)^2\Psi_x\right)_y
  + \left((1-x)^2\Psi_y\right)_x\Biggr) +F(t,x,y)
\label{eq:1}
\end{equation}
for 
$$ 
    0\le x \le 1,\quad -\infty < y < \infty,\quad  t\ge 0 $$
with initial and  boundary conditions
$$
 \Psi(0,x,y)=f(x,y),\quad  \Psi(t,0,y)=q(t,y).  $$
Here $a>0$ and $b$, with $|b|<1$, are real constants and $F$ is a smooth
function.  The outer boundary $\Gamma_1$ at $x=1$ is an ingoing characteristic
so that no boundary condition is allowed. 

Since the boundary data at $\Gamma_0$ can be absorbed into the source $F$, we
treat the case $q=0$ (see the remark in Sec.~\ref{sec:hp}). We denote the
$L_2$ norm over $\Gamma_1$ by
$$
            \|\Psi\|^2{}_{\Gamma_1}=\int dy \Psi^2(t,1,y)
$$
and the $L_2$ norm over the boundary $\Gamma_0$ at $x=0$ by
$$
            \|\Psi\|^2{}_{\Gamma_0}=\int dy \Psi^2(t,0,y).  
$$

We want to show that there is an energy estimate and that the problem is
stable against lower order perturbations. We derive the necessary estimates.
First,
\begin{eqnarray}
  & 2(\Psi,\Psi_{xt})+2a(\Psi,\Psi_t)=-2(\Psi_x,\Psi_t)
 +\partial_t\|\Psi\|^2{}_{\Gamma_1}+a\partial_t\|\Psi\|^2 
\nonumber\\
  &=-\bigg((1-x)\Psi_x,(1-x)\Psi_x\bigg) -\|\Psi_y\|^2
 -2b \bigg((1-x)\Psi_x,(1-x)\Psi_y\bigg) +(\Psi,F) , \nonumber\\
\noalign{\hbox{i.e.}}
 &\partial_t\|\Psi\|^2{}_{\Gamma_1}
  +a\partial_t\|\Psi\|^2+\|(1-x)\Psi_x\|^2+\|\Psi_y\|^2
\nonumber\\
  &+2b\bigg((1-x)\Psi_x,(1-x)\Psi_y\bigg)=2(\Psi_x,\Psi_t) +(\Psi,F).
\label{eq:2}
\end{eqnarray}

Next,
\begin{eqnarray}
 &2(\Psi_t,\Psi_{xt})+2a\|\Psi_t\|^2=\|\Psi_t\|^2{}_{\Gamma_1}
    +2a\|\Psi_t\|^2 \nonumber\\
  &=-\frac{1}{2} \partial_t \bigg(\|(1-x)\Psi_x\|^2+\|\Psi_y\|^2
  +2b((1-x)\Psi_x,(1-x)\Psi_y )\bigg) +(\Psi_t,F).
\label{eq:3}
\end{eqnarray}

Next,
\begin{eqnarray}
 &2(\Psi_x,\Psi_{xt})+2a(\Psi_x,\Psi_t)=\partial_t\|\Psi_x\|^2
   +2a(\Psi_x,\Psi_t)=
  \bigg(\Psi_x,((1-x)^2\Psi_x)_x\bigg)+
   (\Psi_x,\Psi_{yy})& \nonumber\\
   &+ b\bigg(\Psi_x,((1-x)^2\Psi_x)_y\bigg)
   + b\bigg(\Psi_x,((1-x)^2\Psi_y)_x\bigg)+(\Psi_x,F).
\label{eq:4}
\end{eqnarray}
Now,
\begin{eqnarray}\bigg(\Psi_x,((1-x)^2\Psi_x)_x\bigg)&= &-
   \bigg(\Psi_x,2(1-x)\Psi_x \bigg) +\bigg(\Psi_x,(1-x)^2\Psi_{xx}\bigg)
  \nonumber \\
   &=&-(\Psi_x,2(1-x)\Psi_x)-\bigg(((1-x)^2\Psi_x)_x,\Psi_x\bigg)
   -\|\Psi_x\|^2{}_{\Gamma_0} \, , \nonumber
\end{eqnarray}
i.e.
\begin{eqnarray}
\bigg(\Psi_x,((1-x)^2\Psi_x)_x\bigg)
  =-\bigg(\Psi_x,(1-x)\Psi_x\bigg)-{1\over 2}\|\Psi_x\|^2{}_{\Gamma_0}.
   \nonumber
\end{eqnarray}   
Also,
\begin{eqnarray}
   (\Psi_x,\Psi_{yy})&=&-(\Psi_{xy},\Psi_y)
   =-{1\over 2}\|\Psi_y\|^2{}_{\Gamma_1},\nonumber\\
   b\bigg(\Psi_x,((1-x)^2\Psi_x)_y\bigg)
  &=&-b\bigg(((1-x)^2\Psi_x)_y,\Psi_x\bigg)=0,\nonumber\\
  b\bigg(\Psi_x,((1-x)^2\Psi_y)_x\bigg)&=&-2b\bigg(\Psi_x,(1-x)\Psi_y\bigg)
  +b\bigg(\Psi_x,(1-x)^2\Psi_{xy}\bigg)\nonumber\\
  &=&-2b\bigg((1-x)\Psi_x,\Psi_y\bigg). \nonumber
\end{eqnarray}

Therefore (\ref{eq:4}) becomes
\begin{eqnarray}
     \partial_t \|\Psi_x\|^2+\bigg(\Psi_x,(1-x)\Psi_x\bigg)
   +{1\over 2}\|\Psi_x\|^2{}_{\Gamma_0}
 +{1\over2}\|\Psi_y\|^2{}_{\Gamma_1}+2b\bigg((1-x)\Psi_x,\Psi_y\bigg)
  =-2a(\Psi_x,\Psi_t) +(\Psi_x,F).
\label{eq:5}
\end{eqnarray}

All the boundary terms have the right sign to enhance the estimates. Therefore
we ignore them. (See the Remark below.)
Adding the simplified estimates (\ref{eq:2}), (\ref{eq:3}),
(\ref{eq:5})  gives
\begin{eqnarray}
 &\partial_t \bigg(a\|\Psi\|^2+\|\Psi_x\|^2+\frac{1}{2} Q \bigg)
      +Q+\bigg(\Psi_x,(1-x)\Psi_x\bigg)  \nonumber\\
 &=-2b\bigg((1-x)\Psi_x,\Psi_y\bigg)+2(1-a)(\Psi_x,\Psi_t)-2a\|\Psi_t\|^2
  +(\Psi+\Psi_t+\Psi_x,F) \nonumber \\
 &  \le {\rm const.} \bigg( \|\Psi_x\|^2+\|\Psi_y\|^2
   + \|\Psi\|^2+\|F\|^2\bigg),
\label{eq:6}
\end{eqnarray}
where
$$
  Q=\|(1-x)\Psi_x\|^2+\|\Psi_y\|^2+2b\bigg((1-x)\Psi_x,(1-x)\Psi_y\bigg).
$$
Since $|b|<1,$ there is a $\delta>0$ such that
$$ 
Q\ge \delta\left(\|(1-x)\Psi_x\|^2+\|\Psi_y\|^2\right).$$
Therefore (\ref{eq:6}) gives us the required estimate for the energy norm
$$
     E=a\|\Psi\|^2+\|\Psi_x\|^2+\frac{1}{2} Q.
$$. 

\medskip

\noindent Remark: If the boundary term in (\ref{eq:2}) had not been ignored
then we would have obtained a stronger estimate for the energy
$$
     \hat E=E +\|\Psi\|^2{}_{\Gamma_1}
$$
which also controls the growth of $\Psi$ on the boundary $\Gamma_1$.

\medskip

We shall now prove that
the problem is stable against lower order perturbations. We add an expression
$$ 
   P=A\Psi_x+B\Psi_y+C\Psi_t+D\Psi $$
to (\ref{eq:1}). Then the estimates for (\ref{eq:2}), (\ref{eq:3}) and (\ref{eq:4})
will be changed by lower order terms
\begin{eqnarray}
 &(\Psi,A\Psi_x)+(\Psi,B\Psi_y)+(\Psi,C\Psi_t)+(\Psi,D\Psi) \nonumber\\
 &(\Psi_t,A\Psi_x)+(\Psi_t,B\Psi_y)+(\Psi_t,C\Psi_t)+(\Psi_t,D\Psi)\nonumber\\
 &(\Psi_x,A\Psi_x)+(\Psi_x,B\Psi_y)+(\Psi_x,C\Psi_t)+(\Psi_x,D\Psi).\nonumber
\end{eqnarray}
Clearly, there is an energy estimate, provided we choose $2a>|C|$. Thus the
strip problem (\ref{eq:1}) is well posed.

Now we start with
\begin{equation}
   2\Phi_{xt}=\left((1-x)^2\Phi_x\right)_x+\Phi_{yy}
   +b\left((1-x)^2\Phi_x\right)_y
     +b\left((1-x)^2\Phi_y\right)_x + S(t,x,y).
     \label{eq:stph}
\end{equation}
We make the change of variables
$$ 
    \Phi=e^{ax}\Psi,$$
i.e.,
$$
   \Phi_x=e^{ax}\Psi_x+ae^{ax}\Psi,\quad 
   \Phi_{xx}=e^{ax}\Psi_{xx}+2ae^{ax}\Psi_x+a^2e^{ax}\Psi,
$$
and set $F=e^{-ax}S$.
Then we obtain (\ref{eq:1}) which is modified by ${\cal R}$
\begin{eqnarray}
 2(\Psi_x+a \Psi)_t=\left((1-x)^2 \Psi_x\right)_x+ \Psi_{yy}
 +b\left(\left((1-x)^2\Psi_x\right)_y+\left((1-x)^2\Psi_y\right)_x\right)
   +F+{\cal R}.
\label{eq:7}
\end{eqnarray}
Here ${\cal R}$ consists of lower order terms,
$$
{\cal R}=2a(1-x)^2 \Psi_x+\left(a^2(1-x)^2
    -2a(1-x)\right) \Psi+2ab(1-x)^2 \Psi_y.
$$
Since (\ref{eq:1}) is stable against lower order terms there is an energy
estimate for (\ref{eq:7}). In the same way as in Sec.~\ref{sec:hp}  we can
estimate the higher derivatives. This allows us to extend well-posedness to
the variable coefficient problem and, locally in time, to the quasilinear
problem.

\section{The quasilinear wave equation on an asymptotically flat background}
\label{sec:af}

We now treat the null-timelike initial-boundary problem  (\ref{eq:3dwe}) for
the quasilinear wave equation. We compactify the domain $R\le r\le \infty$ by
the transformation $x=1-R/r$ to obtain a strip problem $0\le x \le 1$ with
future null infinity ${\cal I}^+$ at the boundary $x=1$. In terms of the
rescaled field $\hat \Phi =r \Phi$, the wave equation transforms into
\begin{eqnarray}
        &2 \partial_u \partial_x \hat \Phi
    -  R^{-1} \partial_x(W(1-x)^2 \partial_x \hat \Phi) 
        +(1-x) R^{-1}( \partial_x W) \hat \Phi
        \nonumber \\
       & + R^{-2}D_A( (1-x)^2W^A \partial_x \hat \Phi) 
        + R^{-2}\partial_x( (1-x)^2 W^AD_A \hat \Phi)
     - R^{-2}(1-x) (D_AW^A)  \hat \Phi 
        \nonumber \\ 
         &- R^{-1}D_A ( e^{2\beta}D^A \hat \Phi) 
         = - r^3  R^{-1}e^{2\beta} S.
         \label{eq:cphi}
\end{eqnarray}
Here we use the conformal $2$-metric $h_{AB}$ and its inverse $h^{AB}$ to raise and
lower indices of tensor fields on the spacelike $(u= {\rm const.} ,r= {\rm
const.} )$ spherical cross-sections . Up to lower order terms, (\ref{eq:cphi})
is a 3(spatial)-dimensional version of (\ref{eq:stph}) where the
$y$-coordinate has been replaced by the $x^A$-coordinate on the spherical
cross-sections and the $t$-coordinate has been replaced by the
$u$-coordinate.  In order for our treatment  to apply to the quasilinear case,
we assume that the metric coefficients $(W,\beta,W^A,h_{AB})$ depend smoothly
upon $(\Phi,u,r,x^A)$ and that the source $S$ depends smoothly upon 
$(\Phi,\partial_a \Phi,u,r,x^A)$, with non-singular Lorentzian geometry in the
neighborhood of the initial data.

We treat the modified problem resulting from the transformation $\hat \Phi =
e^{ax} \Psi$. The same argument used in Sec.~\ref{sec:2strip} shows that this
problem  is stable with respect to lower order terms. We ignore these terms
and thus obtain the strip problem
\begin{eqnarray}
        &2 \partial_u ( \partial_x \Psi +a \Psi) 
      =  R^{-1}\partial_x(W(1-x)^2 \partial_x \Psi)
       - R^{-2}D_A( (1-x)^2W^A \partial_x \Psi) 
     - R^{-2}\partial_x( (1-x)^2 W^A D_A  \Psi)
         \nonumber \\
        &  + R^{-1}D_A ( e^{2\beta}D^A \Psi)
      +F\, ,\quad  \quad \Psi(0,x,x^A) =f \, ,
        \quad \Psi(u,0,x^A) =q \, ,
         \label{eq:cpsi}
\end{eqnarray}
where $F= - r^3 R^{-1}e^{2\beta} e^{ax}S$. In order to treat (\ref{eq:cpsi}),
we require that the physical space source has asymptotic behavior
$S=O(r^{-3})$ so that $F$ is square integrable over the strip. No boundary
condition is allowed at the outer boundary $\Gamma_1$ at $x=1$ since ${\cal
I}^+$  is an ingoing characteristic surface. 

We obtain the required estimates for (\ref{eq:cpsi}) by the same method used
in Sec.~\ref{sec:2strip}. The data at the inner boundary $\Gamma_0$ at $x=0$
may be absorbed into $F$ so it suffices to treat the case $q=0$. We define the
inner product
$$
       (\Psi_1,\Psi_2)=\int_0^! dx \oint d\omega \Psi_1 \Psi_2
$$
and $L_2$ norm $\|\Psi\|^2 = (\Psi,\Psi)$, where $d\omega$ is the area element
on the unit sphere. We write 
$$
    \|V_A\|^2 = (V_A,V^A)=(h^{AB}V_A, V_B).
$$ 
Since the spherical cross-sections are  spacelike, their intrinsic 2-metric
$h_{AB}$ is positive definite so that  $\|V_A\|$ serves as an $L_2$ norm for
the angular components. We also need a metric norm for spacelike 3-vectors. In
the standard Cauchy problem this is supplied by the intrinsic 3-metric of the
spacelike Cauchy hypersurfaces. Since the characteristic hypersurfaces have a
degenerate 3-metric, we take a different approach. We use the projection
operator $\pi^a_b=\delta^a_b-t^a \partial_b u$, where $t^a\partial_a
=\partial_u$, to define a 3-metric $\gamma^{ab}=\pi^a_c  \pi^b_d g^{cd}$. For
the Bondi-Sachs metric (\ref{eq:nullmet}), the resulting components in the
$(u,r,x^A)$ coordinates are 
$$ 
    \gamma^{au}=0, \quad \gamma^{rr}= e^{-2\beta} W, 
  \quad \gamma^{rA}= -e^{-2\beta} r^{-2}W^A, \quad \gamma^{AB}=r^{-2}h^{AB}.
$$  
Denoting $x^i= (r,X^A)$, this implies that $\gamma^{ij} \rightarrow e^{ij}$ as
$r\rightarrow \infty$, where $e^{ij}$ is the Euclidean 3-metric expressed in
standard spherical coordinates. In the compactified  coordinates, $\tilde
x^i=(x,x^A)$ it is more useful to deal with the rescaled 3-metric $\tilde
\gamma^{ab}= e^{2\beta}r^2 \gamma^{ab}$ which has components
$$ \tilde \gamma^{au}=0, \quad \tilde \gamma^{xx}=(1-x)^2 W, 
  \quad \tilde \gamma^{xA}= -R^{-1}(1-x)^2 W^A,
    \quad \tilde \gamma^{AB}=e^{2\beta}h^{AB}.
$$ 
We then define 
$$
    \|V_i\|^2 = (\tilde \gamma^{ij} V_i, V_j)
$$
which serves as an $L_2$ norm for the $\tilde x^i$ components. Thus
\begin{equation}
          \|\partial_i \Psi\|^2 =
   \|W^{1/2}(1-x)\partial_x\Psi\|^2+\|e^{\beta}D_A \Psi\|^2
     -2R^{-1}\bigg ((1-x)\partial_x\Psi,(1-x)W^A D_A\Psi \bigg ) .
\end{equation}
We also define the corresponding inner products and norms on the boundaries,
e.g.
$$
       (\Psi_1,\Psi_2)_\Gamma= \oint_\Gamma d\omega \Psi_1 \Psi_2 \, , \quad
            \|\Psi\|_\Gamma^2 = (\Psi,\Psi)_\Gamma.
$$

Because the radial coordinate $r$ used in the Bondi-Sachs metric
 (\ref{eq:nullmet}) is a surface area coordinate,
the conformally rescaled $2$-metric $h_{AB}=r^{-2} g_{AB}$ of the spherical
cross-sections has determinant $\det(h_{AB})=\det(q_{AB})$, where 
$q_{AB}$ is the unit sphere metric. Consequently,
\begin{equation}
                 (\Psi_1,D_A D^A \Psi_2) =- (D_A\Psi_1, D^A\Psi_2)
\end{equation} 
and 
\begin{equation}
                 (V^A,D_A \Psi) =- (D_AV^A, \Psi)
\end{equation}
where $V^A(u,x,x^A)$ is  any smooth vector field on the spherical
cross-sections. These identities allow the necessary integration by parts.

We derive the required estimates by freezing the dependence of the metric
coefficients on $(\Psi, u,x)$ but we retain their dependence on $x^A$ so that
$W^A$ and $h_{AB}$ remain smooth vector and tensor fields on the spherical
cross-sections. We follow the procedure in Sec.~\ref{sec:2strip}.  First,
\begin{eqnarray}
  & 2(\Psi,\partial_u \partial_x \Psi)+2a(\Psi,\partial_u\Psi)
  =-2(\partial_x \Psi,\partial_u \Psi)
  +\partial_u\|\Psi\|^2{}_{\Gamma_1} +a\partial_u|\Psi\|^2 
   \nonumber\\
   &=-R^{-1}\bigg(W(1-x)\partial_x \Psi,(1-x)\partial_x \Psi\bigg)   
     -R^{-1}\|e^{\beta}D_A\Psi\|^2
   +2R^{-2} \bigg((1-x)W^A D_A \Psi,(1-x)\partial_x\Psi\bigg) +(\Psi,F) ,   
    \nonumber\\
\noalign{\hbox{i.e.}}
 &\partial_u\|\Psi\|^2{}_{\Gamma_1}+a\partial_u\|\Psi\|^2
       +R^{-1}\|\partial_i \Psi\|^2
 =2(\partial_x \Psi,\partial_u \Psi)  +(\Psi,F).
\label{eq:c2}
\end{eqnarray}

Next,
\begin{eqnarray}
 &2(\partial_u \Psi,\partial_u \partial_x\Psi)+2a\|\partial_u\Psi\|^2
      =\|\partial_u\Psi\|^2{}_{\Gamma_1}+2a\|\partial_u\Psi\|^2 \nonumber\\
  &=-\frac{1}{2} R^{-1}\partial_u \bigg(\|W^{1/2}(1-x)\partial_x \Psi\|^2
   +\|e^{\beta}D_A \Psi\|^2
  -2R^{-1}((1-x)\partial_x\Psi,(1-x)W^A D_A\Psi\bigg)
   +(\partial_u \Psi,F) \, ,  \nonumber
\end{eqnarray}
so that
\begin{equation}
 \|\partial_u\Psi\|^2{}_{\Gamma_1}+2a\|\partial_u\Psi\|^2 
   +\frac{1}{2} R^{-1}\partial_u \|\partial_i \Psi \|^2 
   =(\partial_u \Psi,F).
\label{eq:c3}
\end{equation}

Next,
\begin{eqnarray}
  && 2(\partial_x \Psi,\partial_u\partial_x\Psi)
  +2a(\partial_x\Psi,\partial_u\Psi)
 =\partial_u\|\partial_x\Psi\|^2+2a(\partial_x\Psi,\partial_u\Psi) \nonumber \\
  && =R^{-1}\bigg(\partial_x \Psi,\partial_x(W(1-x)^2\partial_x \Psi)\bigg)
  +R^{-1}\bigg (\partial_x \Psi,D_A  (e^{2\beta}D^A \Psi)\bigg )
 \nonumber\\
    &&-R^{-2}\bigg(\Psi_x, D_A ((1-x)^2W^A \partial_x \Psi)\bigg)
   -R^{-2}\bigg(\partial_x \Psi,\partial_x((1-x)^2 W^A D_A \Psi)\bigg)
     +(\partial_x \Psi,F).
\label{eq:c4}
\end{eqnarray}
As shown in in Sec.~\ref{sec:2strip},
\begin{eqnarray}
  \bigg(\partial_x \Psi,\partial_x(W(1-x)^2 \partial_x\Psi)\bigg)
  =-\bigg(\partial_x\Psi,W(1-x)\partial_x \Psi\bigg)
     -{1\over 2}\|W^{1/2}\partial_x \Psi\|^2{}_{\Gamma_0}.
   \nonumber
\end{eqnarray}   
Also,
\begin{equation}
    (\partial_x\Psi,D_A(e^{2\beta} D^A \Psi))
    =-{1\over 2}\|e^{\beta}D_A\Psi\|^2{}_{\Gamma_1}
\end{equation}
and
\begin{eqnarray}
 \bigg(\partial_x\Psi,D_A((1-x)^2W^A\partial_x \Psi)\bigg)
 + \bigg(\partial_x\Psi,\partial_x((1-x)^2 W^A D_A \Psi)\bigg)
   =-2\bigg((1-x)\partial_x \Psi,W^A D_A \Psi \bigg). \nonumber
\end{eqnarray}
Therefore (\ref{eq:c4}) becomes
\begin{eqnarray}
   &  \partial_u \|\partial_x \Psi\|^2+R^{-1}
     \bigg (\partial_x\Psi,W(1-x)\partial_x \Psi\bigg)
       +{1\over 2}R^{-1}\|W^{1/2}\partial_x \Psi\|^2{}_{\Gamma_0} 
       +{1\over 2}R^{-1}\|e^{\beta}D_A\Psi\|^2{}_{\Gamma_1} \nonumber \\
    & -2R^{-2}\bigg((1-x)\partial_x \Psi,W^A D_A\Psi\bigg)
   =-2a(\partial_x \Psi,\partial_u \Psi) +(\partial_x \Psi,F).
\label{eq:c5}
\end{eqnarray}

As before, the boundary terms have the right sign to enhance the estimates so
that we can ignore them. Adding the simplified estimates (\ref{eq:c2}),
(\ref{eq:c3}), (\ref{eq:c5})  gives
\begin{eqnarray}
  &&\partial_u \bigg(a\|\Psi\|^2+\|\partial_x \Psi\|^2
     +\frac{1}{2} R^{-1}\|\partial_i \Psi \|^2 \bigg)
      +R^{-1}|\partial_i \Psi \|^2
    +R^{-1}\bigg(\partial_x \Psi,W(1-x)\partial_x \Psi\bigg)  \nonumber\\
   &&= 2R^{-2}\bigg((1-x)\partial_x \Psi,W^A D_A\Psi\bigg)
   +2(1-a)(\partial_x \Psi, \partial_u \Psi)-2a\|\partial_u \Psi\|^2
   +(\Psi+\partial_u \Psi+\partial_x \Psi,F) \nonumber \\
 & & \le {\rm const.} \bigg( \|\partial_x \Psi\|^2+\|D_A \Psi\|^2 
    + \|\Psi\|^2 +\|F\|^2\bigg).
\label{eq:c6}
\end{eqnarray}
Therefore (\ref{eq:c6}) gives us an energy estimate provided that the 3-metric
$\gamma^{ij}$ has $(+++)$ signature, so that $\|\partial_i \Psi \|$ is a norm
for the gradient $\partial_i \Psi=(\partial_r \Psi, \partial_A \Psi)$. This is
equivalent to the requirement that the principal part of the wave operator
reduce to an elliptic operator in the stationary case where the $u$-derivatives
vanish.  Since  $\gamma^{ij}$ is asymptotic to the Euclidean metric  as
$r\rightarrow \infty$, this positive-definite condition is satisfied
throughout some exterior domain.

Estimates for the higher derivatives of $\Psi$ and stability against lower
order perturbations follow from the same arguments given in
Sec.~\ref{sec:model}. This establishes the well-posedness of the
worldtube-nullcone problem for the case of smooth variable coefficients. The
extension of well-posedness, locally in time, for the quasilinear case then
follows from the standard techniques referred to in Sec.~\ref{sec:model} .

For a mass  $M$ Schwarzschild geometry, $\gamma^{rr}=e^{-2\beta} W= 1-2M/r$ so
that positive-definiteness of the 3-metric $\gamma^{ij}$
breaks down at $r=2M$ where the worldtube becomes
null. In this limiting case  of the double-null problem, the 
$\partial_u \|\partial_x \Psi\|^2 $ term in (\ref{eq:c6}) suffices to provide the required
estimate. However, for $R<2M$ the ``worldtube'' is spacelike and the  problem
must be treated differently.

\begin{acknowledgments}

J.~W. acknowledges support from NSF grant PHY-0854623 to the University of
Pittsburgh.  

\end{acknowledgments}

\end{document}